\documentclass[sort,comma,authoryear,round,12pt]{elsarticle}

\usepackage{latexsym,amsfonts,amsmath,amssymb}
\usepackage[a4paper]{geometry}
\usepackage{graphicx, color, soul}
\usepackage{subfig, caption}
\usepackage{setspace}
\usepackage{lineno}
\usepackage{bm}
\usepackage{appendix} 
\usepackage{rotating}
\usepackage{ulem}

\usepackage{mathtools}
\newcounter{tableeqn}[table]
\renewcommand{\thetableeqn}{\thetable.\arabic{tableeqn}}
\newcounter{tablesubeqn}[tableeqn]
\renewcommand{\thetablesubeqn}{\thetableeqn\alph{tablesubeqn}}

\newcommand*\patchAmsMathEnvironmentForLineno[1]{%
   \expandafter\let\csname old#1\expandafter\endcsname\csname #1\endcsname
   \expandafter\let\csname oldend#1\expandafter\endcsname\csname end#1\endcsname
   \renewenvironment{#1}%
      {\linenomath\csname old#1\endcsname}%
      {\csname oldend#1\endcsname\endlinenomath}}%
\newcommand*\patchBothAmsMathEnvironmentsForLineno[1]{%
   \patchAmsMathEnvironmentForLineno{#1}%
   \patchAmsMathEnvironmentForLineno{#1*}}%
\AtBeginDocument{%
\patchBothAmsMathEnvironmentsForLineno{equation}%
\patchBothAmsMathEnvironmentsForLineno{align}%
\patchBothAmsMathEnvironmentsForLineno{flalign}%
\patchBothAmsMathEnvironmentsForLineno{alignat}%
\patchBothAmsMathEnvironmentsForLineno{gather}%
\patchBothAmsMathEnvironmentsForLineno{multline}%
}

\journal{Journal of Theoretical Biology}

\begin{document}

\begin{frontmatter}

\title{Enhanced species coexistence in Lotka-Volterra competition models due to nonlocal interactions}

\author{Gabriel Andreguetto Maciel}
\author{Ricardo Martinez-Garcia \corref{cor1}}

\cortext[cor1]{Corresponding author: ricardom@ictp-saifr.org}

\address{ICTP South American Institute for Fundamental Research \& Instituto de F\'isica Te\'orica, Universidade Estadual Paulista - UNESP, Rua Dr. Bento Teobaldo Ferraz 271, Bloco 2 - Barra Funda, 01140-070 S\~ao Paulo, SP, Brazil}

\begin{abstract} 
We introduce and analyze a spatial Lotka-Volterra competition model with local and nonlocal interactions. We study two alternative classes of nonlocal competition that differ in how each species' characteristics determine the range of the nonlocal interactions. In both cases, nonlocal interactions can create spatial patterns of population densities in which highly populated clumps alternate with unpopulated regions. These non-populated regions provide spatial niches for a weaker competitor to establish in the community and persist in conditions in which local models predict competitive exclusion.  Moreover, depending on the balance between local and nonlocal competition intensity, the clumps of the weaker competitor vary from M-like structures with higher densities of individuals accumulating at the edges of each clump to triangular structures with most individuals occupying their centers. These results suggest that long-range competition, through the creation of spatial patterns in population densities,  might be a key driving force behind the rich diversity of species observed in natural ecological communities.
\end{abstract}


\end{frontmatter}

\emph{Declarations of interest:} none \\




\section{Introduction}
The competitive exclusion principle predicts that two species competing for the same resource cannot coexist if environmental factors are constant \citep{Gause1934, Hardin1960}. However, this contrasts with the large number of competitors that often coexist in ecological communities \citep{Hutchinson1961}. Unveiling this paradox and explaining how competing species coexist in nature has been a long-standing goal of ecological theory \citep{May1972, Tilman1982, Chesson, Tokeshi2009}. Some of the mechanisms that have been proposed to explain the coexistence of competing species rely on environmental fluctuations either in space or time \citep{Amarasekare2003, Chesson, Maciel_2018_TE}, diet differentiation \citep{Kartzinel2015}, higher-order or frequency-dependent interactions \citep{Grilli2017,Ayala1971}, or tradeoffs between fitness components \citep{Levins1971,Kneitel2004,Cadotte2006,Angert2009,Tarnita2015a,Martinez-Garcia2017,Martinez-Garcia2020}.

Without any of these ingredients, classical models like Lotka-Volterra predict that two competitors only coexist if individuals from each species compete more strongly with their relatives than with individuals from the other species \citep{Murray1, Chesson}. However, the original Lotka-Volterra model assumes that the populations are well-mixed and two individuals are equally likely to interact with each other regardless of their location \citep{Durrett1994, Lee2001, Hutchinson2007, o2020}. This assumption requires that the interaction's spatial scale is negligible compared to the spatial scales of movement, which is not valid for several biological systems \citep{Hutchinson2007, Holmes1994, Tilman1997, Martinez-Garcia2020a}. Most often, populations are not perfectly mixed, and individuals only interact with neighbors within a small region around them \citep{Lee2001}.

Nonlocal competition (i.e., competition between a focal individual and its neighbors within a finite range) is frequent in natural systems and has been suggested to underlie the emergence of non-uniform spatial distributions of individuals. For example, Martinez-Garcia et al.,\, (\citeyear{Martinez-Garcia,Martinez-Garcia2014,Martinez-Garcia2021}) showed that nonlocal competition alone can create regular vegetation patterns in water-limited ecosystems. In competitive communities, species arrange in regularly spaced clumps on the abstract niche space due to nonlocal competition \citep{Scheffer2006,Pigolotti2007,Fort2009}. Several territorial species and central-place foragers often create a hexagonal, overdispersed pattern of territory packing  and central-place \citep{Tarnita2017,Barlow1974,Peters1975,Doncaster1991}, see also \citet{Pringle2017} for a review. Finally, in more theoretical studies that do not focus on a specific system, pattern formation induced by nonlocal competition has been reported in several extensions of the Fisher-Kolmogorov equation \citep{Fuentes2003, Dornelas2019, Maruvka2006, DaCunha2011}, in the presence of Allee effects \citep{Clerc2005}, and in models for interacting Brownian particles in which reproduction or diffusion rates decay linearly with the population density in a finite neighborhood \citep{Hernandez-Garcia2004, Lopez2004, Lopez2006, Bonachela2012}.

These emergent non-uniform spatial distributions of population densities may substantially impact the dynamics of ecological communities. If spatial patterns in population density segregate species in space, intraspecific interactions become more common than interspecific interactions, which favors species coexistence \citep{Bolker1999, Tilman1997}. \color{black} Also, regions with lower population density can potentially be invaded by a second species to establish a stable coexistence \textit{via} spatial niche partitioning \citep{Chesson}. Previous studies have already covered the effect of spatial patterns and nonlocal interactions on two-species competitive systems \citep{Britton, Segal, Bayliss}. \citet{Britton} studied a reaction-diffusion model with nonlocal competition and local facilitation and derived the conditions for pattern formation, assuming that the intensity of the competition between two individuals decays exponentially with the distance between them. Although this study suggests that species may coexist in situations where local models predict competitive exclusion, it does not provide the conditions required for this to happen and focuses on very restricted combinations of nonlocal competition strengths. Alternative nonlocal extensions of the Lotka-Volterra competition model have been introduced in \citet{Segal} and \citet{Bayliss}. Their analysis, however, is restricted to parameter regimes in which the local model already predicts coexistence. Hence, they do not find evidence of nonlocal interactions reversing species exclusion. \cite{Manna}, study the effect of nonlocal interactions in three-species competitive systems, but only intraspecific interactions are nonlocal. Finally, \citet{Eigentler} and \citet{Eigentler2020} have shown that spatial self-organization can enable coexistence in an ecohydrological model for vegetation dynamics. However, their model accounts for both competition and facilitation and involves spatial tradeoffs because competing species differ in local competition and colonization strength.

In this article, we study a purely competitive Lotka-Volterra model with local and nonlocal interactions and derive analytically the conditions in which nonlocal interactions change the outcome of competition predicted by the model's local equivalent.  We investigate two biologically motivated scenarios differing by how each species' characteristic spatial scale of competition determine the range of nonlocal interactions. In both scenarios, we find that nonlocal interactions allow species to coexist for a broad range of parameter combinations for which the local model predicts competitive exclusion. This result suggests that reversal of competitive exclusion is a rather general property of nonlocal interactions with the potential to occur in several natural systems.

\section{Methods: a Lotka-Volterra model with nonlocal competition} \label{the_model}

We study a spatial version of the two-species competitive Lotka-Volterra equations in a one-dimensional domain. In addition to exponential growth and local competition, our model also includes individual dispersal and nonlocal interactions, two inherently spatial processes. The density of individuals of each species thus changes according to
\begin{align}
  \frac{\partial \rho_1(x,t)}{\partial t}  = & \ D_1 \frac{\partial^2 \rho_1}{\partial x^2} \ + b_1\,\rho_1 \left(1 - \frac{a_{11}\rho_1 + a_{12}\rho_2}{K_1} - h_{11} \widetilde{\rho}_{11} - h_{12} \widetilde{\rho}_{12} \right) \label{eq:lvnl1} \\
  \frac{\partial \rho_2(x,t)}{\partial t}  = & \ D_2 \frac{\partial^2 \rho_2}{\partial x^2} \ + b_2\,\rho_2 \left(1 - \frac{a_{21}\rho_1 + a_{22}\rho_2}{K_2} - h_{21} \widetilde{\rho}_{21} - h_{22} \widetilde{\rho}_{22} \right), \label{eq:lvnl2}
\end{align}
where we have dropped the dependence in $x$ and $t$ from the density fields in the right side of the equations to simplify the notation. The first term in each equation accounts for dispersal with species-specific diffusion $D_i$ ($i = 1, 2$). The second term corresponds to species intrinsic growth at rate $b_i$ and pairwise local and nonlocal competition. Local competition is modeled as in the classical Lotka-Volterra model with competition coefficients $a_{ij}$ ($i, j = 1, 2$). The local interactions represent, for example, competition for space or for any other resource that is gathered locally by the individuals. $K_i$ is the carrying capacity of species $i$. Hence, we fix the coefficients for intraspecific competition $a_{ii}=1$ to ensure that $K_i$ is one of the stationary values of the population density in one-species populations without nonlocal interactions. The coefficients $h_{ij}$ give the intensity of nonlocal competition interactions of species $i$ with itself and with the other competitor. $\widetilde{\rho}_{ij}$ is the average density of $j$-individuals in a neighborhood centered at $x$, which is a proxy for the mean number of $i$-$j$ interactions within that neighborhood. In general, we can weigh the average by the effect that the $i$-$j$ inter-individual distance has on the intensity of the interaction between them \citep{Britton, Lee2001}. If we assume spatial isotropy, $\widetilde{\rho}_{ij}$ is:
\begin{equation}
\widetilde{\rho}_{ij}(x,t) = \int dx'\,G_{ij}(\rvert x-x'\rvert ) \rho_j(x',t)
\end{equation}
where $G_{ij}$ is a kernel function that weighs the influence that individuals from species $j$ exert on the growth rate of species $i$ when they are at a distance $\rvert x-x' \rvert $ \citep{Murray2002V2}. Because $\widetilde{\rho}_{ij}$ are average densities, all the interaction kernels that we consider are normalized. Non-normalized kernels would only add a constant to $\tilde{\rho}_{ij}$ that can be absorbed by the corresponding $h_{ij}$ parameter without affecting our results qualitatively. These nonlocal terms represent, for example, competition for resources that are acquired by individuals within a finite neighborhood or direct interference mediated by forces that have a non-zero range of action \citep{Schenk2006, Granato2019}. For example, in plant communities, the nonlocal terms could account for root-mediated competition for resources. Local competition could represent competition for space during plant establishment or competition for light, assuming that the size of the canopy is negligible compared to the range of the root system.

\begin{figure}[!h]
    \centering \includegraphics[width=0.8\textwidth]{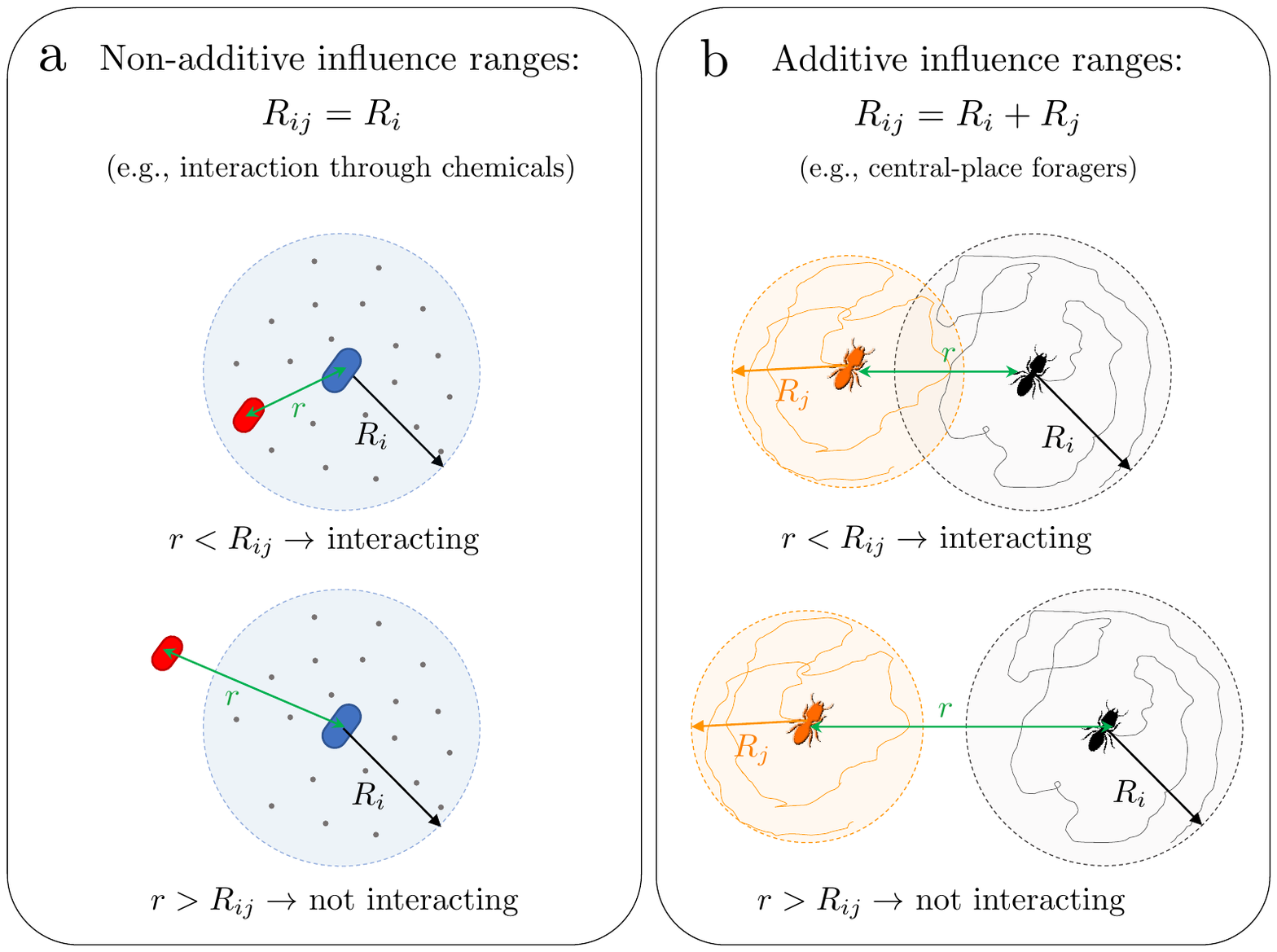} 
    \caption{Schematic representation of a $i$-$j$ interspecific interaction for non-additive (a), and additive species influence ranges (b). a) Competitive interactions mediated by the release of chemicals, such as toxins (small dots in the figure), show non-additive species influence ranges. Individuals from species $j$ (red ellipses) interact with individuals of the species $i$ (blue ellipses) if they lie within the reach of the $i$-cell secretions (blue circles limited by dashed lines). b) Central-place foragers, such as termites \citep{Tarnita2017}, are an example of non-local competition with additive influence ranges. Two individuals interact if their foraging territories overlap. In the figure, the circles represent the foraging territories of the two individuals, $r$ is the distance between the centers of the foraging territories, and the erratic curves within each of the territories represent two foraging trajectories.} \label{fig:scheme}
\end{figure}

For all the analyses that follow, we consider normalized top-hat kernel functions for the sake of mathematical simplicity. This is equivalent to assuming that individuals interact with each other with the same intensity if the distance between them is below a threshold value. We refer to this threshold value setting the maximum distance at which two individuals interact with each other as the interaction range. For two species, $i$ and $j$, with interaction range $R_{ij}$, the kernel function is given by:
\begin{equation}
G_{ij}(r) = \Pi_{2R_{ij}}(r) =
\begin{cases}
\frac{1}{2 R_{ij}}  \ \ \ \text{if} \ \ \  r \leq R_{ij} \\
\ 0 \  \  \ \  \text{if} \ \ \  r > R_{ij}.
\end{cases} \label{eq:step_function}
\end{equation}
The interaction range $R_{ij}$ is, in general, a function of the species-specific ranges of competition, $R_i$ and $R_j$. These spatial scales, $R_i$ and $R_j$, define the range over which the presence of individuals from one of the species, $i$ or $j$, respectively, decreases the growth rate of other individuals. The formal definition of the interaction range $R_{ij}$, however, will depend on the nature of the long-range interaction that dominates the system under study. We will work with two possible definitions for the interaction range, depending on how it emerges from the species-specific spatial scales of competition $R_i$ and $R_j$: 
\begin{itemize}
\item[-] \textit{Non-additive influence ranges (scenario I):} the interaction range is equal to the influence range of one of the species. In this case, individuals of species $i$ are affected by conspecifics if the inter-individual distance is less or equal to the influence range of species $i$, $r\leq R_i$. The kernel is $G_{ii}(r) = \Pi_{2R_i}(r)$, where $\Pi_{2R_i}$ is the top-hat function of Eq.\,(\ref{eq:step_function}) with $R_{ij} = R_i$. Similarly, $G_{jj} = \Pi_{2R_j}(r)$, and $G_{ij}(r) = \Pi_{2R_j}(r)$ and $G_{ji}(r) = \Pi_{2R_i}(r)$ for interspecific interactions. An instantiation of this scenario is, for example, a competitive interaction between species with the ability to release toxins (Fig.\,\ref{fig:scheme}a) \citep{Bais2003, Granato2019}.

\item[-]  \textit{Additive influence ranges (scenario II):} the interaction range is equal to the sum of the influence ranges of both species. In this scenario, the range of interspecific competition is $R_i+R_j$,  and the ranges of the intraspecific competitions are $2R_i$ and $2R_j$ for species $i$ and $j$, respectively.  In terms of the kernels, this leads to $G_{ij} = G_{ji}(r) = \Pi_{2(R_i+R_j)}(r)$, $G_{ii}(r) = \Pi_{4R_i}(r)$, and $G_{jj} = \Pi_{4R_j}(r)$ respectively. An example of this scenario is the competition for a shared pool of resources. If resources are located at $x$, they can be foraged by $i$--individuals within a region $[x-R_i, x+R_i]$ and by $j$--individuals within a region $[x-R_j, x+R_j]$. The effective range of competition is thus the addition of both influence ranges and a similar rationale follows for instraspecific competitions (Fig.\,\ref{fig:scheme}b).
\end{itemize}

In the next sections, we study the long-term behavior of the model for non-additive influence ranges and discuss the most important differences that appear when influence ranges are additive (we provide the full analysis for this case in \ref{app:scenarioII}). We combine numerical and analytical methods and focus on understanding how nonlocal competition and the emergent spatial patterns of population densities stabilize species coexistence. We performed numerical simulations using the method of lines in a spatial grid of size $L = 10$ and periodic boundary conditions (\cite{Schiesser}; see SM for further details).

\section{Results} \label{results}

To reduce the number of free parameters we first derive the nondimensional version of Eqs.\,(\ref{eq:lvnl1})-(\ref{eq:lvnl2}) for non-additive influence ranges. We use $b_1^{-1}$ as the reference time scale and $R_1$ as the reference length scale. Therefore, the nondimensional influence range of species 1 is the unity and the nondimensional influence range of species 2 is given by the ratio $q = R_2/R_1$. A similar argument follows for the rescaled time, and we write the new space and time variables as $\xi = x/R_1$ and $\tau = b_1 t,$ respectively. Using the nondimensional quantities $\ u_1 = \rho_1/K_1, \ \ u_2 = \rho_2/K_2, \ d_1 = D_1/(d_1 R_1^2), \ d_2 = D_2/(d_1 R_1^2), \ c_{12} = a_{12} K_2/K_1, \ c_{21} = a_{21} K_1/K_2, \ \beta = b_2/b_1, \ \phi_{11}= K_1 h_{11}, \ \phi_{22} = K_2 h_{22}, \ \phi_{12} = K_2 h_{12}, \ \phi_{21} = K_1 h_{21}$ in Eqs.\,(\ref{eq:lvnl1})-(\ref{eq:lvnl2}), the model equations become:
\begin{align}
  \frac{\partial u_1}{\partial \tau} = & \ d_1 \frac{\partial^2 u_1}{\partial \xi^2} + u_1 \bigg[1 - u_1 - c_{12} u_2 \notag \\
                                   &   - \phi_{11} \int \Pi_{2}(|\xi - \xi'|) u_1(\xi', \tau) d\xi' -  \phi_{12} \int \Pi_{2q}(|\xi - \xi'|) u_2(\xi', \tau) d\xi' \bigg] \label{eq:non_dim_1} \\  
  \frac{\partial u_2}{\partial \tau} =& \ d_2 \frac{\partial^2 u_2}{\partial \xi^2} + \beta u_2 \bigg[1 - u_2 - c_{21} u_1 \notag \\
  &   - \phi_{22} \int \Pi_{2q}(|\xi - \xi'|) u_2(\xi', \tau) d\xi' - \phi_{21} \int \Pi_{2}(|\xi - \xi'|) u_1(\xi', \tau) d\xi' \bigg]. \label{eq:non_dim_2}
\end{align}

\subsection{Stability of the spatially uniform steady states}

Because all the interaction kernels are normalized, the uniform steady states $u^*_1$ and $\ u^*_2$ are the same for non-additive and additive influence ranges [Eqs.\,(\ref{eq:non_dim_1})-(\ref{eq:non_dim_2}) and \ref{app:scenarioII}]. We find four uniform steady states. The first one represents the extinction of both species, the second and the third, the competitive exclusion of each of the species, and  the fourth one, the coexistence of both species. The values for  $u^*_1$ and $u^*_2$ in each of these cases are provided in Table\,\ref{Table1}. To determine the linear stability of these states we write perturbed solutions around the uniform steady states, $u_1(\xi, \tau) = u^*_1 + \epsilon v_1(\xi, \tau)$ and $u_2(\xi, \tau) = u^*_2 + \epsilon v_2(\xi, \tau)$ with $\epsilon\ll 1$, and study the short-time growth rate of the perturbations $v_i$. 

\begin{table}[htb] 
\centering
\stepcounter{table}
\def\arraystretch{1}
\begin{tabular}{llr}
 \hline
 Fixed point & Population densities & \\
    \hline \\
    Trivial 					& $\displaystyle u^*_1 = 0, \ u^*_2 = 0$ & \refstepcounter{tableeqn} (\thetableeqn)\label{eq:tab1-1} \\ [3ex]
    Species 1 - only 	& $\displaystyle u^*_1 = \frac{1}{1 + \phi_{11}}$, \ $u^*_2 = 0$ & \refstepcounter{tableeqn} (\thetableeqn)\label{eq:tab1-2} \\ [3ex]
    Species 2 - only 	& $\displaystyle u^*_1 = 0$, \ $\displaystyle u^*_2 = \frac{1}{1 + \phi_{22}}$ & \refstepcounter{tableeqn} (\thetableeqn)\label{eq:tab1-3} \\ [3ex]
    Coexistence 		& $\displaystyle u^*_1 = \frac{q_2 - p_2}{D}$, $\displaystyle u^*_2 = \frac{p_1 - q_1}{D}$ & \refstepcounter{tableeqn} (\thetableeqn)\label{eq:tab1-4} \\ [2ex]  \hline
    \end{tabular} 
    \addtocounter{table}{-1}
    \caption{Spatially uniform steady states of systems of Eqs.\,(\ref{eq:non_dim_1})-(\ref{eq:non_dim_2}) and (\ref{eq:scenarioII_eq1})-(\ref{eq:scenarioII_eq2}). Here $p_1= 1 + \phi_{11}$, $p_2 = c_{12} + \phi_{12}$, $q_1 = c_{21} + \phi_{21},$ $q_2 = 1 + \phi_{22}$ and $D = p_1 q_2 - p_2 q_1$.}\label{Table1}
\end{table}

First, we study the local limit of the model in which the kernel functions are Dirac delta functions and our model converges to the classical competitive Lotka-Volterra equations. In this limit,  the scenarios with additive and non-additive influence ranges are equivalent to each other and the model equations reduce to:
\begin{align}
  \frac{\partial u_1}{\partial \tau} =& \, d_1 \frac{\partial^2 u_1}{\partial \xi^2} + u_1 \left[1 - (1+\phi_{11}) u_1 - (c_{12} + \phi_{12}) u_2\right] \label{eq:local1} \\  
  \frac{\partial u_2}{\partial \tau} =& \, d_2 \frac{\partial^2 u_2}{\partial \xi^2} + \beta u_2 \left[1 - \left(1 + \phi_{22} \right) u_2 - \left(c_{21}+ \phi_{21} \right) u_1\right]. \label{eq:local2}
\end{align}

\noindent Hence, if all competitive interactions are purely local or equivalently if perturbations to the steady states of the nonlocal model are spatially uniform, the stability conditions of the homogeneous steady states are the well-known stability conditions of the Lotka-Volterra competition model. Competitive exclusion occurs if the interspecific competition is stronger than the intraspecific competition. If this condition is met for only one of the species, the only outcome of the model is the competitive exclusion of the weaker competitor. If it holds for both species, however, the system is in a bistable regime of mutual exclusion in which the identity of the excluded species depends on initial population densities. Finally, if the intraspecific competition is stronger than the interspecific competition for both species, the model predicts their stable coexistence. Therefore, like in the classical Lotka-Volterra competitive equations, both species coexist only if the interspecific interaction is weak \citep{Murray1}. The mathematical relations between parameters that reflect these stability conditions are summarized in Table \,\ref{tab:stability_local}.

\begin{table}[htb]
\centering
\stepcounter{table}
\def\arraystretch{1}
\begin{tabular}{llr}
 \hline
 Fixed point & Stability condition & \\
    \hline \\
    Trivial & Always unstable \\ [2ex]
    Species 1 - only & $\displaystyle \phi_{21} + c_{21} > 1 + \phi_{11} $ & \refstepcounter{tableeqn} (\thetableeqn)\label{eq:tab2-1}\\ [2ex]
    Species 2 - only & $\displaystyle \phi_{12} + c_{12} > 1 + \phi_{22}$ & \refstepcounter{tableeqn} (\thetableeqn)\label{eq:tab2-2}\\ [2ex]
    Coexistence      & $\displaystyle 1 + \phi_{11} > \phi_{21} + c_{21} $ and $1 + \phi_{22} > \phi_{12} + c_{12}$ & \refstepcounter{tableeqn} (\thetableeqn)\label{eq:tab2-3}\\ [1ex] \hline
  \end{tabular}
      \addtocounter{table}{-1}
    \caption{Stability conditions of the uniform steady states for the equivalent local model. When only local interactions are considered scenarios I and II are equivalent.}
  \label{tab:stability_local}
\end{table}

Next, we study the full, nonlocal model. Substituting the perturbed solutions in Eqs.\,(\ref{eq:non_dim_1})-(\ref{eq:non_dim_2}) and retaining only linear terms in $\epsilon$, we obtain the following system of equations for the perturbations $v_1(\xi, \tau)$ and $v_2(\xi, \tau)$:
\begin{align}
  \frac{\partial v_1(\xi, \tau)}{\partial \tau} =& \ d_1 \frac{\partial^2 v_1}{\partial \xi^2} + \left[ 1 - u^*_1 - (1 + \phi_{11}) u^*_1 - (c_{12} + \phi_{12}) u^*_2 \right] v_1 - c_{12} u^*_1 v_2   \notag \\
                             &    - u^*_1 \left(\phi_{11} \int \Pi_{2}(|\xi - \xi'|) v_1(\xi', \tau) d\xi' +  \phi_{12} \int \Pi_{2q}(|\xi - \xi'|) v_2(\xi', \tau) d\xi' \right) \label{eq:perturb_eq1} \\  
  \frac{\partial v_2(\xi, \tau)}{\partial \tau} =& \ d_2 \frac{\partial^2 v_2}{\partial \xi^2} + \beta \left[ 1 - u^*_2 - (1 + \phi_{22}) u^*_2 - (c_{21} + \phi_{21}) u^*_1 \right] v_2 - \beta c_{21} u^*_2 v_1 \notag \\
                                &    - \beta u^*_2 \left(\phi_{22} \int \Pi_{2q}(|\xi - \xi'|) v_2(\xi', \tau) d\xi' +  \phi_{21} \int \Pi_{2}(|\xi - \xi'|) v_1(\xi', \tau) d\xi' \right). \label{eq:perturb_eq2}  
\end{align}

\noindent Because Eqs.\,(\ref{eq:perturb_eq1})-(\ref{eq:perturb_eq2}) are linear integro-differential equations with constant coefficients,  we can solve them applying the Fourier transform to the spatial coordinate. Written in matrix form, the Fourier transformed system of equations is
\begin{equation}
\frac{\partial \hat{\textbf{v}}(k,t)}{\partial \tau} = A \hat{\textbf{v}}(k,\tau),
\end{equation}
\noindent where $\hat{\textbf{v}}(k,\tau)$ is the vector containing the Fourier transform of each perturbation, $\hat{\textbf{v}}(k,\tau) = [\hat{v}_1(k,\tau) \  \hat{v}_2(k,\tau)]^T$ and $k$ is the wavenumber of each mode. $A$ is a $2 \times 2$ matrix with elements
\begin{align}
  A_{11} =& \,1 - u^*_1 - p_1 u^*_1 - p_2 u^*_2 - d_1 k^2 - \phi_{11} u^*_1 \widehat{\Pi}_{2} (k) \\
  A_{12} =& - \left[c_{12} + \phi_{12} \widehat{\Pi}_{2q} (k)\right] \, u^*_1 \\        
  A_{21} =& - \beta \left[c_{21} + \phi_{21} \widehat{\Pi}_{2} (k)\right] \, u^*_2 \\        
  A_{22} =& \,  \beta - \beta u^*_2 - \beta q_2 u^*_2 - \beta q_1 u^*_1 - d_2 k^2 - \beta \phi_{22} u^*_2 \widehat{\Pi}_{2q} (k), 
\end{align}
\noindent where we have used the convolution theorem to transform the convolution integrals and the parameters $p_1,$ $p_2$, $q_1$ and $q_2$ are combinations of the original model parameters as defined in Table \ref{Table1}. A stationary state $(u^*_1,$ $u^*_2)$ is stable if all the modes of the perturbation decay to zero with time, or equivalently, if the amplitude of the perturbation decays with time for any wavenumber. This means that the real part of all the eigenvalues of $A$ must be negative, or, according to the Routh-Hurwitz criterion, $\text{Tr}(A) < 0$ and $\det(A) > 0$. Imposing these two conditions to $A$, we obtain the two stability conditions for each stationary state:
\begin{align}
  & - d_1 k^2 + 1 - p_1 u^*_1 - p_2 u^*_2 - (1 + \phi_{11} \widehat{\Pi}_{2} (k)) u^*_1 \notag \\
  &     \quad \quad    - d_2 k^2 + \beta - \beta q_2 u^*_2 - \beta q_1 u^*_1 - \beta (1 + \phi_{22} \widehat{\Pi}_{2q} (k)) u^*_2 < 0 \\
  \nonumber \\
  & \left[- d_1 k^2 + 1 - p_1 u^*_1 - p_2 u^*_2 - (1 + \phi_{11} \widehat{\Pi}_{2} (k)) u^*_1\right]\Big[- d_2 k^2 + \beta - \beta q_2 u^*_2 - \beta q_1 u^*_1 \notag \\
  &   \quad  \quad  - \beta (1 + \phi_{22} \widehat{\Pi}_{2q} (k))u^*_2\Big] - \beta u^*_1 u^*_2(c_{12} + \phi_{12} \widehat{\Pi}_{2q} (k))(c_{21} + \phi_{21} \widehat{\Pi}_{2} (k)) > 0 . 
\end{align}

\noindent These conditions can be particularized for each of the four pairs of values $(u_1^*, \,  u_2^*)$ provided in Table~\ref{Table1} to obtain the stability conditions for the uniform steady states with nonlocal competitive interactions.  We provide the stability conditions in Table \ref{tab:stability_scenarioI}. 

\begin{table}[!h]
\centering
\stepcounter{table}
\def\arraystretch{1}
\begin{tabular}{clr}
 \hline
 Fixed point & Stability condition & \\
    \hline \\
    Trivial & Always unstable \\ [3ex]
    Species 1 & $\displaystyle \left(1 + \phi_{11}\right) d_2 k^2 + \beta \left[\phi_{21} + c_{21} - \left(1 + \phi_{11}\right) \right] > $ & \\
    only & $   \quad  \quad - \left[\left(1 + \phi_{11}\right) d_1 k^2 + 1 + \phi_{11} \widehat{\Pi}_{2} (k)\right]$ & \stepcounter{tableeqn}\refstepcounter{tablesubeqn}(\thetablesubeqn) \label{eq:tab3-1a}\\  [1.5ex]
    & $\displaystyle \left[(1 + \phi_{11}) d_1 k^2 + 1 + \phi_{11} \widehat{\Pi}_{2} (k)\right]$ \\ 
    & $       \quad  \quad  \times \left[(1 + \phi_{11}) d_2 k^2 + \beta \left(\phi_{21} + c_{21} - (1 + \phi_{11}) \right) \right] > 0$  & \refstepcounter{tablesubeqn}(\thetablesubeqn) \label{eq:tab3-1b} \\ [4ex]
    Species 2 & $(1 + \phi_{22}) d_1 k^2 + \phi_{12} + c_{12} - (1 + \phi_{22}) > $ \\
      only   & $  \quad \quad  - \left[(1 + \phi_{22}) d_2 k^2 + \beta \left(1 + \phi_{22} \widehat{\Pi}_{2q} (k)\right)\right]$ & \stepcounter{tableeqn}\refstepcounter{tablesubeqn}(\thetablesubeqn) \label{eq:tab3-2a} \\ [1.5ex]
    & $\displaystyle \left[(1 + \phi_{22}) d_2 k^2 + \beta\left(1 + \phi_{22} \widehat{\Pi}_{2q} (k)\right)\right] $ \\
    & $\displaystyle  \quad \quad   \times \left[(1 + \phi_{22}) d_1 k^2 + \phi_{12} + c_{12} - (1 + \phi_{22})\right] > 0$ &\refstepcounter{tablesubeqn}(\thetablesubeqn) \label{eq:tab3-2b} \\ [4ex]
    Coexistence & $ (d_1 + d_2) k^2 + u^*_1 \left(1 + \phi_{11} \widehat{\Pi}_{2}(k)\right) + \beta u^*_2 \left( 1 + \phi_{22} \widehat{\Pi}_{2q}(k) \right) > 0$ & \stepcounter{tableeqn}\refstepcounter{tablesubeqn}(\thetablesubeqn) \label{eq:tab3-3a} \\ [1.5ex]
                  & $\displaystyle u^*_1 u^*_2 \left[\left(\frac{d_1 k^2}{u^*_1} + 1 + \phi_{11} \widehat{\Pi}_{2}(k)\right)\left( \frac{d_2 k^2}{u^*_2} + \beta \left(1 + \phi_{22} \widehat{\Pi}_{2q}(k) \right) \right) \right.$ \\  [2.5ex]
                   &   $\quad \quad  \displaystyle  \left.       - \beta \left(c_{12} + \phi_{12} \widehat{\Pi}_{2q}(k)\right)\left(c_{21} + \phi_{21} \widehat{\Pi}_{2}(k)\right)\right] > 0$ & \refstepcounter{tablesubeqn}(\thetablesubeqn) \label{eq:tab3-3b} \\ [2ex]
    \hline
  \end{tabular}
    \addtocounter{table}{-1}
    \caption{Stability conditions for the uniform steady states of with non-additive influence ranges (scenario I). The steady states are given in Table \ref{Table1}.}
\label{tab:stability_scenarioI}
\end{table}

\subsection{Conditions for population pattern formation and ecological consequences}

The nonlocal interactions can destabilize the stationary states for parameter values in which the local model is stable. In these cases, some of the modes of the perturbation grow and lead to periodic spatial distributions of population  densities. Because we are mainly interested in studying how long-range interactions change the stability of the steady states of competitive Lotka-Volterra models, we first set $d_1=d_2=0$ and study the effect of diffusion in our results later. 

\subsubsection{Non-additive species influence ranges: scenario I}\label{subsec:scenI}

\begin{figure}[!h]
    \centering \includegraphics[width=0.9\textwidth]{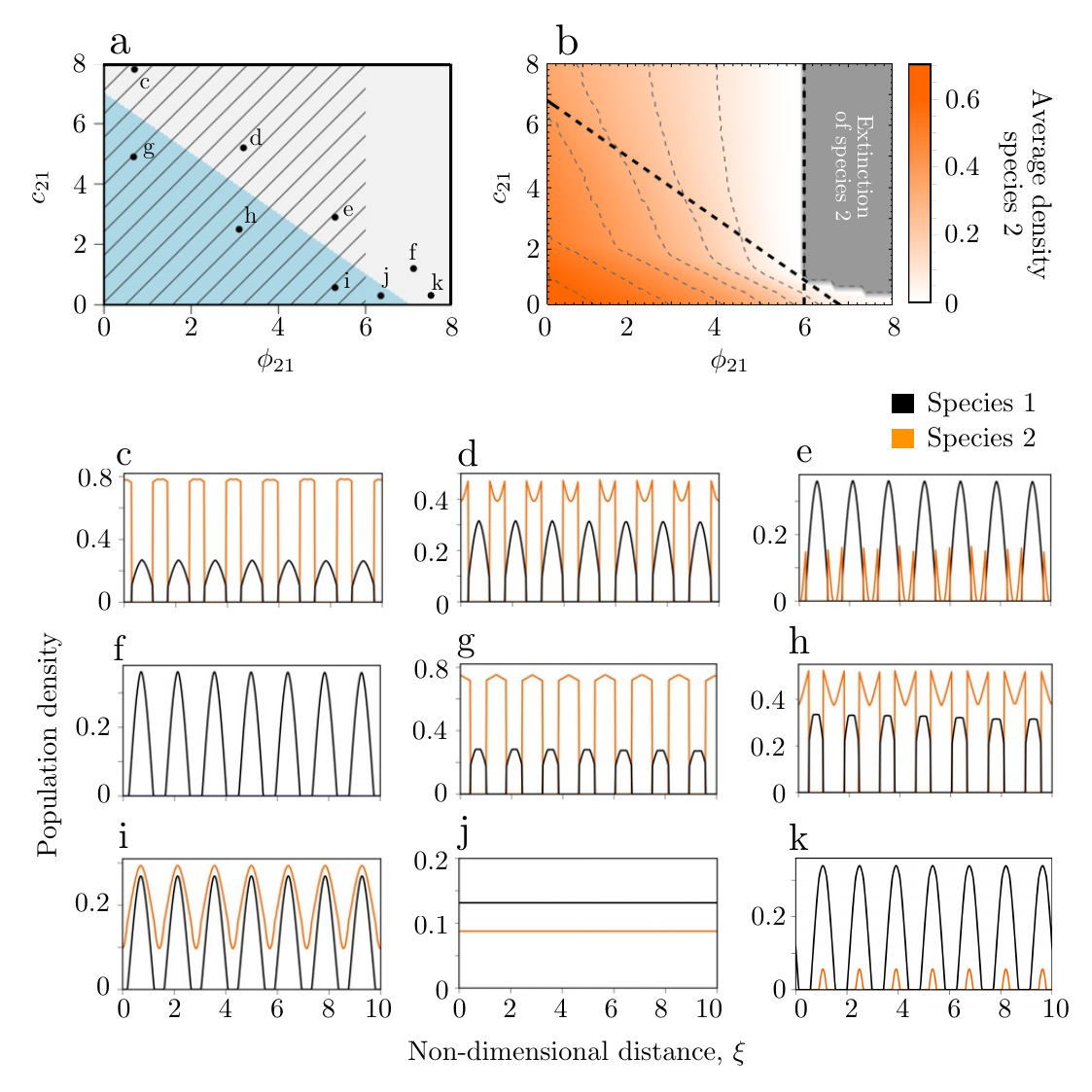} 
    \caption{a) Parameter regimes for species exclusion and coexistence. The light blue and gray regions indicate the conditions in which the local model predicts species coexistence and exclusion respectively. The diagonal stripes highlight where inequality (\ref{eq:inv_condition3}) is satisfied and hence the nonlocal model predicts species coexistence. The dots indicate the pairs $c_{21}$, $\phi_{21}$ used to obtain the spatial patterns shown in panels (c)-(k). b) Average density of species 2 as a function of $c_{21}$ and $\phi_{21}$, $\tilde{u}_2\neq 0$ indicates species coexistence and $\tilde{u}_2 = 0$ competitive exclusion of species 2. The black dashed lines indicate the end of the striped and light blue region in panel a. c-k) Long-time spatial patterns of population density. Each panel corresponds to pairs $c_{21}$, $\phi_{21}$ as labeled in panel a. Other parameters are: $c_{12} = 0.4$, $\beta = 1$, $\phi_{11} = 6,$ $\phi_{12} = 0.5$ and $\phi_{22} = 0.5$.} \label{bifur1}
\end{figure}

Because we are interested in studying whether nonlocal interactions change the outcome of the competitive interaction as compared to the local model, we focus on the steady states in which the local model predicts either coexistence or the exclusion of species 2. The case in which species 1 is excluded can be analyzed following the same steps. We parameterize the model such that individuals of species 2 exert a stronger competition on conspecifics than on individuals from species 1, $1 + \phi_{22} >  \phi_{12} + c_{12}$. Given this relationship between model parameters, if $1 + \phi_{11} < \phi_{21} + c_{21}$ the local model predicts the exclusion of species 2 [light gray region in Fig.\,\ref{bifur1}(a)], and stable coexistence otherwise [light blue region in Fig.\,\ref{bifur1}(a)]. \\

\textit{Exclusion of species 2 in the local model: $1 + \phi_{11} < \phi_{21} + c_{21}$.} First, we analyze the case in which the local model predicts the competitive exclusion of species 2. Moreover, we set the intensity of the nonlocal intraspecific competition for species 1 such that it develops self-organized spatial patterns in the mono-species steady state. From Eq.\,(\ref{eq:tab3-1b}) in Table \ref{tab:stability_scenarioI} with $d_1=d_2=0$, this requires that
\begin{equation}\label{eq:sta-excl2}
1 + \phi_{11} \widehat{\Pi}_{2} (k) < 0.
\end{equation}
Because $\phi_{11}$ is a positive parameter, a necessary condition for pattern formation is that the Fourier transform of the kernel for intraspecific competition becomes negative for some wave number \citep{Martinez-Garcia, Pigolotti2007, Dornelas2019}. This condition is met by the top-hat kernels defined in Eq.\,(\ref{eq:step_function}), whose $1D$ Fourier transform is the sinc function. Once the kernel function is conveniently chosen so that patterns are possible, the intensity of the nonlocal intraspecific competition for species 1 needs to be sufficiently large to make competition less intense within a clump than in the unpopulated areas between population clumps.  If we denote this critical value by $\phi_{11}^*$, then $\phi_{11}^* =\,\rvert \widehat{\Pi}_2(k_c)\rvert\approx 4.6$ where $k_c$ is the wavenumber at which the Fourier transform of the kernel function has its absolute minimum. For the rest of our analysis, we fix $\phi_{11}>\phi_{11}^*$ and species 1 will always self-organize and develop spatial patterns upon exclusion of species 2.

Self-organized patterns in the population density of species $1$ may leave regions of the space in which the density of individuals is very low (Fig.\,\ref{fig:SMminu1}). Next, we investigate whether, and in which conditions, these regions create opportunities for species 2 to invade a resident population of species 1 and establish a stable coexistence of both species. To this end, we consider a general solution for the spatial distribution of species 1, $u_1 = U_1(\xi)$, with a zero population of species 2 and derive analytical approximations for the conditions in which a small population of species 2 grows by linearizing the model around this unknown state.

We write the perturbed solution near the state $(U_1(\xi), 0)$ as $u_1(\xi, \tau) = U_1(\xi) + \epsilon\, v_1(\xi, \tau)$ and $u_2(\xi, \tau) = \epsilon \, v_2(\xi, \tau)$. Inserting these solutions in Eq.\,(\ref{eq:non_dim_2}) and retaining only linear terms in $\epsilon$, we get the following equation for $v_2$:
\begin{equation}
  \frac{\partial v_2}{\partial \tau} = \beta \left[1 - c_{21} U_1(\xi) - \phi_{21} \int \Pi_{2}(|\xi - \xi'|) U_1(\xi') d\xi' \right] v_2,
\end{equation}
 from where we can obtain the condition for the invasion of species 2,
\begin{equation}
  1 - c_{21} U_1(\xi) - \phi_{21} \int \Pi_{2}(|\xi - \xi'|) U_1(\xi') d\xi' > 0. \label{eq:inv_condition1}
\end{equation}

\noindent From Eq.\,(\ref{eq:inv_condition1}), it follows that to invade the low populated regions left by the spatial self-organization of species 1, species 2 must overcome local and non-local competition. To further simplify the condition for invasion, we obtain from Eq.\,(\ref{eq:non_dim_1}) with $d_1=0$ an expression for the integral term in Eq.\,(\ref{eq:inv_condition1}) when species 1 shows spatial patterns and species 2 is excluded,
\begin{equation}\label{eq:app-integral}
\int \Pi_{2}(|\xi - \xi'|) U_1(\xi') d\xi' = \frac{1 - U_1(\xi)}{\phi_{11}}.
\end{equation} 

\noindent Inserting Eq.\,(\ref{eq:app-integral}) in Eq.\,(\ref{eq:inv_condition1}) we obtain
\begin{equation}
    \left(\phi_{11} - \phi_{21}\right) + \left(\phi_{21} - c_{21} \phi_{11} \right) U_1(\xi) > 0, \label{eq:inv_condition2}
\end{equation}

\noindent that still depends on the spatial distribution of species 1, $U_1(\xi)$, which is unknown. However, 
as can be seen in SM Fig.\,\ref{fig:SMminu1}, in the regions between the maxima of the patterned distribution of species 1 $U_1(\xi) \approx 0$ for most of the parameter space, which leads to a simple sufficient condition for species 2 to invade,
\begin{equation}
  \phi_{11} > \phi_{21}, \label{eq:inv_condition3}
\end{equation}
\noindent which means that the non-local effects of species 1 must be  greater on itself than on species 2. 

In Figure \ref{bifur1}, we have indicated with diagonal stripes the region in which inequality (\ref{eq:inv_condition3}) is satisfied and species 2 can potentially invade species 1. Notice that because the condition in (\ref{eq:inv_condition3}) does not depend on $c_{21}$, the region with diagonal stripes extends to the light gray region in Fig.~\ref{bifur1} where species coexistence is already possible without nonlocal interactions. To test our analytical approximation, we obtained the steady-state spatial distributions of population densities numerically. First, we sampled the $c_{21}-\phi_{21}$ parameter space and obtained the spatial average of the density of species 2 in the stationary state, $\tilde{u}_2$. Because our parameter choice does not allow for the exclusion of species 1, $\tilde{u}_2 > 0$ indicates species coexistence, and $\tilde{u}_2 = 0$ exclusion of species 2 [Fig.\,\ref{bifur1}(b)]. Second, for some of the sampled values of $c_{21}$  and $\phi_{21}$, we show the spatial patterns of population densities, which confirms that the nonlocal model has a larger region of species coexistence because self-organized spatial patterns in the strong competitor (species 1) create low populated regions in which the weak competitor (species 2) can grow. 

The specific spatial distribution of the populations depends on the intensity of the local and nonlocal competition, $c_{21}$ and $\phi_{21}$ respectively. If $\phi_{21}$ is small and $c_{21}$ is large, species 2 grows in the regions left empty by the spatial pattern of species 1, reaching a maximum population density close to the center of the regions unpopulated by species 1 [Fig.\,\ref{bifur1}(c)]. As $\phi_{21}$ increases and $c_{21}$ decreases, species 2 adopts a M-shape pattern, redistributing individuals from the center of the regions empty of species 1 to the edges [Fig.\,\ref{bifur1}(d)]. If we continue to increase $\phi_{21}$ and decrease $c_{21}$, species 2 fully accumulates at the edges of regions left empty by species 1 because individuals establishing in the center of these areas interact with two clumps of the patterned distribution of species 1, while species-2 individuals close to the edges minimize the intensity of the nonlocal competition with species 1 [Fig.\,\ref{bifur1}(e)]. Finally, when $\phi_{21}>\phi_{11}$ the relation in (\ref{eq:inv_condition3}) is violated, species  2 can no longer persist and the stationary state consists of a spatial pattern of species 1 [Fig.\,\ref{bifur1}(f)].

\textit{Species coexistence in the local model: $1 + \phi_{11} > \phi_{21} + c_{21}$.} Long-range interactions do not change the outcome of species competition in the parameter regime in which the local model predicts stable species coexistence. However, because the condition for species 2 invading low-populated regions left by a patterned distribution of species 1 is partially met when the local model predicts coexistence [see stripped diagonal pattern overlapping partially with the light blue region in Fig.\,\ref{bifur1}(a)], we find two mechanisms of coexistence when nonlocal interactions are included in the model. First, for weak nonlocal and strong local interspecific competition (small $\phi_{21}$ and large $c_{21}$) the population density of species 2 accumulates in the center of the regions that are empty of species 1 [Fig.\,\ref{bifur1}(g)]. As $\phi_{21}$ increases and $c_{21}$ decreases,  the population density within the clumps of species 2 redistributes towards the edges of the regions in-between clumps of species 1. As a consequence, the clumps change their shape and adopt an M-like structure  [Fig.\,\ref{bifur1}(h)]. When $c_{21}$ is much smaller than $\phi_{21}$, species 2 grows in the clumps of species 1 because the nonlocal contribution to the interspecific competition is much stronger than the local one [Fig.\,\ref{bifur1}(i)]. In this parameter regime, species may also coexist without forming spatial patterns [Fig.\,\ref{bifur1}(j)].

\textit{Species coexistence not predicted by analytical calculations.} Finally, our numerical analysis reveals an additional route towards species coexistence 
that is not predicted by our analytical calculations.  Specifically, when $c_{21}$ is small, invasion by species 2 is possible because the second term on the left side of the invasion condition (\ref{eq:inv_condition2}) is sufficiently positive. In this case, species 2 grows in regions with a high density of species 1. However, because $U_1(\xi)$ is always smaller than one,  this mechanism for species coexistence only operates when $c_{21}$ is very small [Fig.\,\ref{bifur1}(k)].
  
\subsubsection{Additive species influence ranges: scenario II}

In many aspects, the results for additive influence ranges are very similar to the results obtained for the non-additive case. Hence, we provide a detailed analysis in \ref{app:scenarioII} and only discuss here the most salient differences between both situations. Following the same rationale of section \ref{subsec:scenI}, we study the non-diffusion limit and focus on a parameter regime in which the local model predicts the competitive exclusion of species 2 and a uniform spatial distribution of species 1. Our goal is, again, to determine the conditions in which nonlocal interactions may change the outcome of the competitive interaction, possibly via the arrangement of species 1 in a self-organized spatial pattern. The stability conditions of the state with a uniform distribution of species 1 and exclusion of species 2 are the same obtained for scenario I [see Eqs. (\ref{eq:tab4-1a})-(\ref{eq:tab4-1b}) in Table \ref{tab:stability_scenarioII} of \ref{app:scenarioII}]. The only difference between both scenarios is the set of wavenumbers that are unstable and can potentially grow once the uniform distribution of species 1 loses stability and spatial patterns form. However, the ecological implications of these emergent patterns can be different in scenario II as compared to scenario I. To study this difference between both scenarios we repeat the invasibility analysis of section \ref{subsec:scenI} now considering additive influence ranges. The invasion condition for species 2 is now given by
\begin{equation}
  1 - c_{21} V_1(\xi) - \phi_{21} \int \Pi_{2(1 + q)}(|\xi - \xi'|) V_1(\xi') d\xi' > 0, \label{eq:inv_condition_scenarioII}
\end{equation}
\noindent where $V_1(\xi)$ is the spatial distribution of species 1 in the absence of species 2 that can be obtained from the steady-state solution of Eq.\,(\ref{eq:scenarioII_eq1}). However, note that the convolution integrals in (\ref{eq:inv_condition_scenarioII}) and Eq.\,(\ref{eq:scenarioII_eq1}) are different and we cannot, in general, repeat the steps followed in section \ref{subsec:scenI} to simplify the invasibility conditions. The only exception is the limit case $q=1$ in which 
(\ref{eq:inv_condition_scenarioII}) and (\ref{eq:scenarioII_eq1}) have the same kernel $\Pi_{2(1+q)}(|\xi - \xi'|)$  and the conclusions from section \ref{subsec:scenI} are valid if we take an effective interaction range equal to two instead of unity. 

\begin{figure}
    \centering
    \includegraphics[width= 0.8\textwidth]{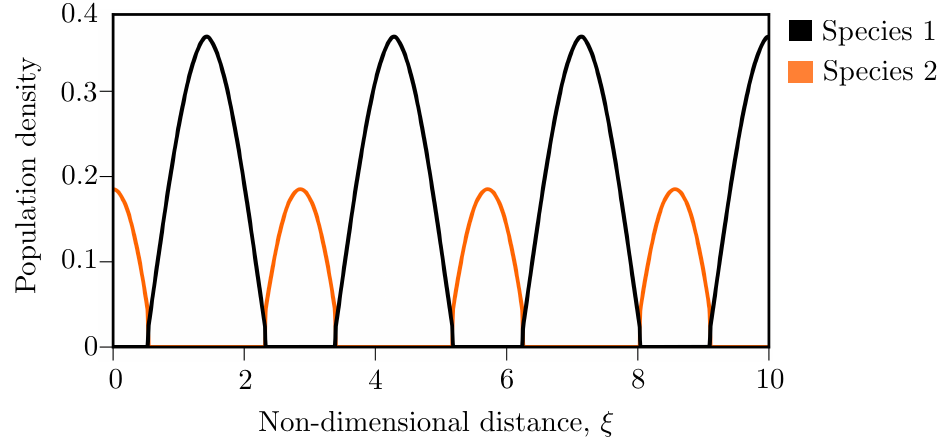}
    \caption{Stationary spatial patterns of species density for additive influence ranges [scenario II, Eqs.\,(\ref{eq:scenarioII_eq1})-(\ref{eq:scenarioII_eq2})]. Interspecific competition coefficients are the same as those in Fig.\,\ref{bifur1}(f): $c_{21} = 1.2$ and $\phi_{21} = 7.1$. Other parameters are: $q=0.2$, $c_{12} = 0.4$, $\beta = 1$, $\phi_{11} = 6$, $\phi_{12} = 0.5$ and $\phi_{22} = 0.5$.}\label{fig:scenario_II}
\end{figure}

In the most general case, the invasion of species 2 depends on its influence range $q \neq 1$. Because $V_1(\xi)$ does not depend on $q$, the condition in (\ref{eq:inv_condition_scenarioII}) indicates that $q< 1$ favors the invasion of species 2.  In Figure \ref{fig:scenario_II}, we show the asymptotic spatial distribution of population densities when $q=0.2$ and all other parameters take the same values used in Fig. \ref{bifur1}(f) for non-additive influence ranges.  Species coexistence is possible when influence ranges are additive, while species 2 is not able to invade, regardless of its influence range, if the interaction is non-additive [Fig.\,\ref{bifur1}(f)]. Conversely, when $q > 1$, the integration interval in condition (\ref{eq:inv_condition_scenarioII}) increases and makes the invasion of species 2 harder. Note, however, that because the convolution integral in (\ref{eq:inv_condition_scenarioII}) returns the average of $V_1$ in a spatial window of length $2(1 + q)$, the negative effect of larger values of $q$ on the invasion of species 2 does not grow unboundedly with $q$. For example, for points at the center of a region between two peaks of species 1, the convolution integral reaches a maximum when the integration interval contains two neighbor maxima of $u_1$.

\begin{figure}
    \centering
       \includegraphics[width= 0.85\textwidth]{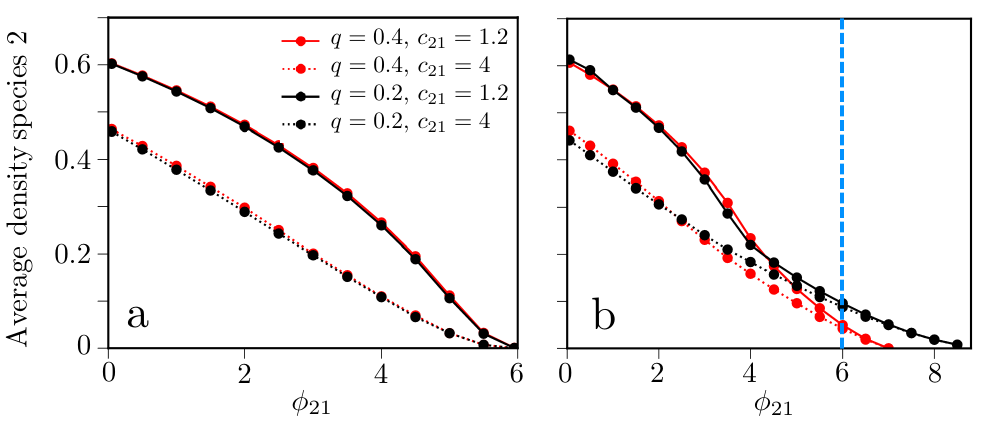}
    \caption{Stationary average densities of species 2 as a function of the nonlocal interspecific competition coefficient $\phi_{21}$. Other parameters are as in Figure \ref{bifur1}: $c_{12} = 0.4$, $\beta = 1$, $\phi_{11} = 6,$ $\phi_{12} = 0.5$ and $\phi_{22} = 0.5$. Panel a corresponds to non-additive influence ranges (scenario I) and panel b, to additive influence ranges (scenario II). The vertical blue dashed line in panel b indicates the critical value of $\phi_{21}$ for the survival of species 2 for non-additive influence ranges, $\phi_{21} = \phi_{11}$.} \label{bifur2}
\end{figure}

Finally, we study the response of species 2 to increased nonlocal and local interspecific competition $\phi_{21}$ and $c_{21}$ when species have different influence ranges and both for non-additive and additive influence ranges [Fig.\,\ref{bifur2}(a) and (b), respectively]. In either scenario (non-additive or additive influence ranges), species 2 decreases when $\phi_{21}$ increases and is eventually eliminated. For non-additive influence ranges, the average density of species 2 does not depend on $q$ (Fig.\,\ref{bifur2}a). For additive influence ranges, however, the density of species 2 is only independent of $q$ for small $\phi_{21}$. When $\phi_{21}$ increases, the effect of $q$ on the density of species 2 increases too and may determine the survival of the population at very large $\phi_{21}$. Finally, the effect of $c_{21}$ has the opposite effect on the average population density of species 2 than $q$: it is negligible for high $\phi_{21}$ and increases when the nonlocal interspecific competition decreases (Fig.\,\ref{bifur2}b).

\subsection{The effect of diffusion}

In this section, we test the robustness of pattern-induced species coexistence against non-zero diffusion rates. To this end, we focus on non-additive influence ranges (scenario I), Eqs.\,(\ref{eq:non_dim_1})-(\ref{eq:non_dim_2}), with $q=1$ and two different combinations of competition coefficients, corresponding to those used in Fig.\,\ref{bifur1}(c) and \ref{bifur1}(d). For each of these two sets of competition coefficients, we vary $d_1$ with $d_2=0$, $d_2$ with $d_1=0$, and $d_1$ and $d_2$ simultaneously with $d_1 = d_2$. 

\begin{figure}[htb]
    \centering
    \includegraphics[width=0.9\textwidth]{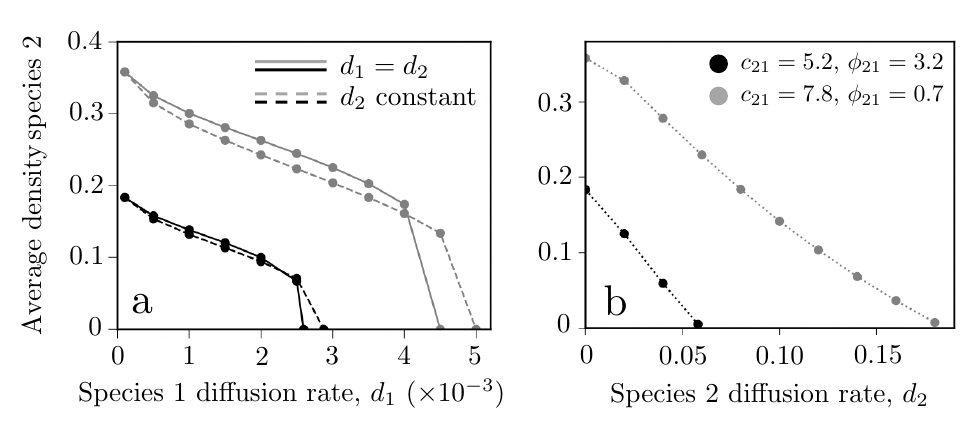}
    \caption{Effect of diffusion on species coexistence for non-additive influence ranges [scenario I, Eqs.\,(\ref{eq:non_dim_1})-(\ref{eq:non_dim_2})]. a) Stationary-state average density of species 2  as function of species 1 diffusion rate, $d_1$. $d_2 = d_1$ for solid curves and $d_2 = 0$ for dashed curves. b) Stationary-state average density of species 2 versus $d_2$ with $d_1 = 0$. The color code is the same in both panels as shown in the top right corner of (b). Grey and black curves have the same competition coefficients that those for Fig.\,\ref{bifur1}(c) and \ref{bifur1}(d) respectively.} \label{bifur_diffusion}
\end{figure}

 As we saw in previous sections, for this set of parameters, the local model predicts the competitive exclusion of species 2 and the nonlocal model the stable coexistence of both species if diffusion rates are equal to zero. In general, diffusion tends to reduce the density of species 2, eventually causing its extinction. Moreover, diffusion of species 1, $d_1$ has a stronger effect on species coexistence than $d_2$ [note the different scales in the abscissa axes of Fig.\,\ref{bifur_diffusion}(a) and \ref{bifur_diffusion}(b)]. This is because while increasing $d_2$ increases the visits of species 2 into regions where species 1 is abundant, $d_1$ acts by disrupting the spatial pattern of species 1, which is a key condition to reverse the competitive exclusion predicted by the local model. Recall that in the absence of species 2, spatial patterns of species 1 develop if $(1 + \phi_{11}) d_1 k^2 + 1 + \phi_{11} \widehat{\Pi}_{2} (k)$ is negative for some wavenumber $k$. Finally, because additive and non-additive influence ranges lead to similar species competition outcomes when $q=1$, we expect only quantitative changes in these results for additive influence ranges.

\section{Discussion}
\label{discussion}

We introduced a Lotka-Volterra competition model with local and nonlocal terms and explored the consequences of the nonlocal terms on the competitive dynamics. We defined and analyzed two different scenarios depending on how the range of the nonlocal interaction depends on each species' influence ranges. In the first scenario, which we termed \textit{non-additive influence ranges}, individuals affect competitors of the same or other species in a spatial range that depends only on the influence range of the individual itself. This scenario could represent, for example, allelopathic interactions between plants in which the extension of the roots or the branches determine species' influence range \citep{Bertin2003, Bais2003, Cheng2015}. Another example could be the different types of antagonistic interactions established in microbial communities and governed by the release of chemicals to the environment \citep{Czaran2002, Nadell2016, Granato2019}. In the second scenario, which we called \textit{additive influence ranges}, the ranges of the nonlocal interactions are a combination of the influence ranges of the species engaged in the interaction. This situation corresponds, for example, to plants competing for resources through their roots or home-ranging animals competing for forage over a finite region, such as central place foragers \citep{Houston1985, Olsson2008}. In these cases, influence ranges are related to root extension and maximum foraging distances, respectively.   

Our results show that nonlocal interactions can create a wide variety of spatial patterns in population density and that those patterns can have significant long-term consequences for the community. Specifically, these spatial patterns can stabilize the community and make species coexist under conditions in which models without nonlocal terms predict competitive exclusion. This reversal of competitive exclusion occurs because the weaker competitor establishes in regions where, due to the stronger competitor's non-homogeneous distribution, interspecific interactions are weak. Spatial niche partitioning is a well-known general mechanism for species coexistence \citep{Chesson, Albrecht2001, Winder2009, Schuette2013} and our study provides a general, analytically tractable theoretical framework to obtain the conditions in which spatial niche partitioning leads to species coexistence in homogeneous environments.  

For non-additive influence ranges, we identified three conditions for competitive exclusion reversal: (i) small diffusion, (ii) spatial pattern of the stronger competitor with nonpopulated regions between clumps, and (iii) the stronger competitor exerts a more intense nonlocal effect on the weaker competitor than on itself, $\phi_{21} > \phi_{11}$. These same conditions apply for additive influence ranges when both species have the same influence range ($q = 1$). In this case, however, results depend strongly on the ratio between influence ranges, and species may coexist even when condition (iii) is not satisfied. In either scenario, additive or non-additive influence ranges, diffusion hinders the reversal of competitive exclusion because its main effect is to break the spatial patterns needed for species coexistence. Our results are also robust against different choices of the competition kernels, provided they meet the conditions for pattern formation \citep{Pigolotti2007, Martinez-Garcia, Martinez-Garcia2014} and the patterns they cause leave some regions empty of the strong competitor \citep{Dornelas2019}.
   
Our work extends previous studies in many ways. First, we extended the analysis in \citet{Britton} and showed that spatial self-organization is a mechanism for species coexistence even in purely competitive ecosystems, without invoking any positive interaction between species. Moreover, we derived the conditions for species coexistence and the emergent spatial patterns of population densities, which allowed us to obtain a much deeper understanding of the mechanisms that allow species coexistence. Finally, \citet{Britton} limits to the case in which interaction kernels are species-specific, which corresponds to setting $G_{11} = G_{21}$ and $G_{22} = G_{12}$ in our model equations. By relaxing this assumption, our framework allows a more general analysis of how multiple spatial scales can come into play and how they may interact with each other to determine the spatial distribution of the populations and the stability of the community. Other competitive models accounting for nonlocal competition have also restricted to the case in which kernels are the same for all interactions ($G_{11} = G_{12} = G_{21} = G_{22}$) or considered non-additive influence ranges (scenario I in this article; $G_{11} = G_{21}$ and $G_{12} = G_{22}$). Besides, those studies focus on a parameter regime in which species already coexist in the equivalent local models and do not provide evidence of nonlocal interactions reversing competitive exclusion \citep{Segal, Bayliss}.

\section*{Acknowledgments}

GAM was supported by a Postdoctoral Fellowship from S\~ao Paulo Research Foundation (FAPESP), grant 2019/21227-0. RMG was supported by FAPESP through Programa Jovens Pesquisadores em Centros Emergentes 2019/24433-0 and 2019/05523-8, Instituto Serrapilheira through grant Serra-1911-31200, and the Simons Foundation. GAM and RMG are supported by FAPESP through grant ICTP-SAIFR 2016/01343-7. This research was supported by resources supplied by the Center for Scientific Computing (NCC/GridUNESP) of the S\~ao Paulo State University (UNESP).

\section*{References}
\bibliography{references_nonlocal} 
 
 \newpage

\appendix
\section{Stability and invasibility conditions for additive influence ranges (scenario II)}\label{app:scenarioII}

The equations for additive influence ranges are:
\begin{align}
  \frac{\partial u_1}{\partial \tau} =& d_1 \frac{\partial^2 u_1}{\partial \xi^2} + u_1 \bigg[1 - u_1 - c_{12} u_2 \notag \\
                                   & \ - \phi_{11} \int \Pi_{4}(|\xi - \xi'|) u_1(\xi', \tau) d\xi' - \phi_{12} \int \Pi_{2(1 + q)}(|\xi - \xi'|) u_2(\xi', \tau) d\xi' \bigg] \label{eq:scenarioII_eq1} \\  
  \frac{\partial u_2}{\partial \tau} =& d_2 \frac{\partial^2 u_2}{\partial \xi^2} + \beta u_2 \bigg[1 - u_2 - c_{21} u_1 \notag \\
  & \ - \phi_{22} \int \Pi_{4q}(|\xi - \xi'|) u_2(\xi', \tau) d\xi' - \phi_{21} \int \Pi_{2(1 + q)}(|\xi - \xi'|) u_1(\xi', \tau) d\xi' \bigg]. \label{eq:scenarioII_eq2}
\end{align}

\subsection{Stability of uniform steady states}

To investigate the stability of the uniform steady states $(u^*_1, u^*_2)$ we write the perturbed solutions around these states as $u_1(\xi, \tau) = u^*_1 + \epsilon v_1(\xi, \tau)$ and $u_2(\xi, \tau) = u^*_2 + \epsilon v_2(\xi, \tau),$ where $\epsilon \ll 1.$ Recall that the uniform steady states are the same as for scenario I and are given in Table\,\ref{Table1} in the main text. Next, we insert the expressions for the perturbed uniform steady states into Eqs.\,(\ref{eq:scenarioII_eq1}) and (\ref{eq:scenarioII_eq2}) and only retain linear terms in $\epsilon$. The linearized equations for the perturbations are:

\begin{align}
  \frac{\partial v_1(\xi, \tau)}{\partial \tau} & = d_1 \frac{\partial^2 v_1}{\partial \xi^2} + \left[1 - u^*_1 - (1 + \phi_{11}) u^*_1 - (c_{12} + \phi_{12}) u^*_2\right] v_1 - c_{12} u^*_1 v_2   \notag \\
                                    - & u^*_1 \left(\phi_{11} \int \Pi_{4}(|\xi - \xi'|) v_1(\xi', \tau) d\xi' +  \phi_{12} \int \Pi_{2(1 + q)}(|\xi - \xi'|) v_2(\xi', \tau) d\xi' \right) \label{eq:perturb_scenarioII_eq1} \\  
  \frac{\partial v_2(\xi, \tau)}{\partial \tau} & = d_2 \frac{\partial^2 v_2}{\partial \xi^2} + \beta \left[1 - u^*_2 - (1 + \phi_{22}) u^*_2 - (c_{21} + \phi_{21}) u^*_1\right] v_2 - \beta c_{21} u^*_2 v_1 \notag \\
                                    - & \beta u^*_2 \left(\phi_{22} \int \Pi_{4q}(|\xi - \xi'|) v_2(\xi', \tau) d\xi' +  \phi_{21} \int \Pi_{2(1 + q)}(|\xi - \xi'|) v_1(\xi', \tau) d\xi' \right). \label{eq:perturb_scenarioII_eq2}  
\end{align}

\noindent Following the same arguments given in the main text for non-additive influence ranges, this system of linear integro-differential equations can be solved using the Fourier transform. By Fourier transforming Eqs.\,(\ref{eq:perturb_scenarioII_eq1}) and (\ref{eq:perturb_scenarioII_eq2}) and applying the convolution theorem we obtain a linear system of equations for the amplitude of the different modes $\hat{v}_1(k,\tau)$ and $\hat{v}_2(k,\tau),$ where $k$ is the wave number. In matrix form, the transformed system is:
\begin{equation}
\frac{\partial \hat{\textbf{v}}(k,\tau)}{\partial \tau} = A \hat{\textbf{v}}(k,\tau),
\end{equation}
\noindent where $\hat{\textbf{v}} = [\hat{v}_1(k,\tau) \ \hat{v}_2(k,\tau)]^T$ as in the main text and $A$ is a $2\times2$ matrix with elements
\begin{align}
  A_{11} =& \,1 - u^*_1 - p_1 u^*_1 - p_2 u^*_2 - d_1 k^2 - \phi_{11} u^*_1 \widehat{\Pi}_{4} (k) \\
  A_{12} =& - (c_{12} + \phi_{12} \widehat{\Pi}_{2(1 + q)} (k)) \, u^*_1 \\        
  A_{21} =& - \beta (c_{21} + \phi_{21} \widehat{\Pi}_{2(1 + q)} (k)) \, u^*_2 \\        
  A_{22} =& \,  \beta - \beta u^*_2 - \beta q_2 u^*_2 - \beta q_1 u^*_1 - d_2 k^2 - \beta \phi_{22} u^*_2 \widehat{\Pi}_{4q} (k). 
\end{align}

\noindent Finally, we use the Routh-Hurwitz Criterion ($\text{Tr}(A) < 0$ and $\det(A) > 0$) to obtain the stability conditions shown in Table \ref{tab:stability_scenarioII}.

\begin{table}[!h]
\centering
\stepcounter{table}
\def\arraystretch{1}
\begin{tabular}{clr}
 \hline
 Fixed point & Stability condition & \\
    \hline
    Trivial & Always unstable \\ [4ex]
    Species 1 & $(1 + \phi_{11}) d_2 k^2 + \beta \left[\phi_{21} + c_{21} - (1 + \phi_{11}) \right] > $ \\
    only & $ \quad \quad - \left[(1 + \phi_{11}) d_1 k^2 + 1 + \phi_{11} \widehat{\Pi}_{4} (k)\right]$ & \stepcounter{tableeqn}\refstepcounter{tablesubeqn}(\thetablesubeqn) \label{eq:tab4-1a}\\  [1.5ex]
    & $\left[(1 + \phi_{11}) d_1 k^2 + 1 + \phi_{11} \widehat{\Pi}_{4} (k)\right]$ \\
    & $  \quad \quad   \times \left[(1 + \phi_{11}) d_2 k^2 + \beta \left(\phi_{21} + c_{21} - (1 + \phi_{11}) \right)\right] > 0$ & \refstepcounter{tablesubeqn}(\thetablesubeqn) \label{eq:tab4-1b} \\ [4ex]
    Species 2& $(1 + \phi_{22}) d_1 k^2 + \phi_{12} + c_{12} - (1 + \phi_{22}) > $ \\
    only     & $ \quad \quad  - \left[(1 + \phi_{22}) d_2 k^2 + \beta\left(1 + \phi_{22} \widehat{\Pi}_{4q} (k)\right)\right]$ & \stepcounter{tableeqn}\refstepcounter{tablesubeqn}(\thetablesubeqn) \label{eq:tab4-2a}\\ [3ex] 
    & $\left[(1 + \phi_{22}) d_2 k^2 + \beta\left(1 + \phi_{22} \widehat{\Pi}_{4q} (k)\right)\right] $ \\
    & $  \quad \quad  \times [(1 + \phi_{22}) d_1 k^2 + \phi_{12} + c_{12} - (1 + \phi_{22})] > 0$ & \refstepcounter{tablesubeqn}(\thetablesubeqn) \label{eq:tab4-2b} \\ [4ex]
    Coexistence & $ (d_1 + d_2) k^2 + u^*_1 \left(1 + \phi_{11} \widehat{\Pi}_{4}(k)\right) + \beta u^*_2 \left(1 + \phi_{22} \widehat{\Pi}_{4q}(k)\right) > 0$ & \stepcounter{tableeqn}\refstepcounter{tablesubeqn}(\thetablesubeqn) \label{eq:tab4-3a}\\  [1.5ex]
                  & $\displaystyle u^*_1 u^*_2 \left[\left(\frac{d_1 k^2}{u^*_1} + 1 + \phi_{11} \widehat{\Pi}_{4}(k)\right)\left(\frac{d_2 k^2}{u^*_2} + \beta \left(1 + \phi_{22} \widehat{\Pi}_{4q}(k)\right)\right)\right.$ \\ [3ex]
                   & $\displaystyle \left.  \quad \quad  - \beta \left(c_{12} + \phi_{12} \widehat{\Pi}_{2(1 + q)}(k)\right)\left(c_{21} + \phi_{21} \widehat{\Pi}_{2(1 + q)}(k)\right)\right] > 0$ & \refstepcounter{tablesubeqn}(\thetablesubeqn) \label{eq:tab4-3b} \\ [2ex]
    \hline
  \end{tabular}
      \addtocounter{table}{-4}
    \caption{Stability conditions for the uniform steady states. The steady states are given in Table \ref{Table1} in the main text.}
  \label{tab:stability_scenarioII}
\end{table}

\subsection{Invasibility of a non-uniform spatial distribution}

In this section we investigate the conditions for the invasion of species 2 of a resident species 1 distributed according to this non-uniform state.We denote by $V_1(\xi)$ the asymptotic spatial distribution of species 1 in the absence of species 2 for $d_1 = 0$ and write perturbed solutions around the state $(V_1(\xi), 0)$ as $u_1(\xi, \tau) = V_1(\xi) + \epsilon v_1(\xi, \tau)$ and $u_2(\xi, \tau) = \epsilon v_2(\xi, \tau)$. Inserting these perturbed solutions in Eqs. (\ref{eq:perturb_scenarioII_eq1}) and (\ref{eq:perturb_scenarioII_eq2}) and retaining only linear terms in $\epsilon$,  the growth of $v_2(\xi, \tau)$ is given by:
\begin{equation}
  \frac{\partial v_2}{\partial \tau} = \beta \left(1 - c_{21} V_1(\xi) - \phi_{21} \int \Pi_{2 (1 + q)}(|\xi - \xi'|) V_1(\xi') d\xi' \right) v_2.
\end{equation}
\noindent Hence, the invasion of species 2 is determined by condition (\ref{eq:inv_condition_scenarioII}) given in the main text.

\newpage
\section{Supplementary Figures}

\begin{figure}[!h]
    \centering
    \includegraphics[width= 0.6\textwidth]{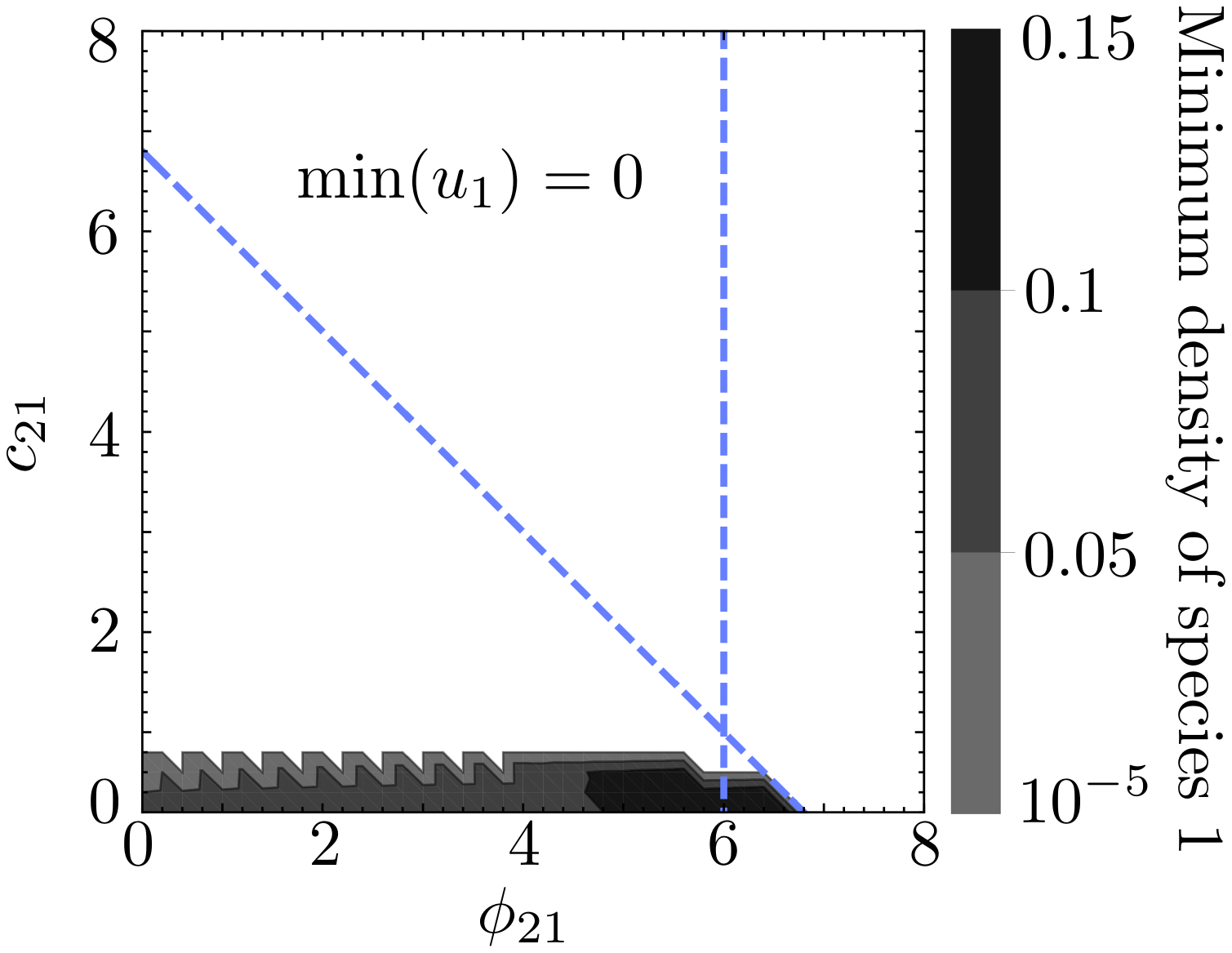}
    \caption{Minimum density of species $1$ as function of $c_{21}$ and $\phi_{21}$. The blue dashed lines indicate the end of the striped and light blue region in Fig.\,\ref{bifur1}(a).  Other parameters are like in Fig.\,\ref{bifur1}: $c_{12} = 0.4$, $\beta = 1$, $\phi_{11} = 6,$ $\phi_{12} = 0.5$ and $\phi_{22} = 0.5$. }\label{fig:SMminu1}
\end{figure}

\newpage
\section{Details of numerical calculations}\label{app:numerics}

All integro-differential equations in this work were solved using the method of lines \citep{Schiesser}. This method uses a semidiscretization approach in which only space is initially discretized and the problem becomes one of solving a system of coupled ordinary differential equations. We can then take advantage of highly sophisticated and accurate methods and packages available for solving ODEs. Here we used Python odeint numerical integrator.

Initial conditions were taken as perturbations on the single species steady states $u^*_1(x) = 1/(1 + \phi_{11})$ and $u^*_2(x) = 1/(1 + \phi_{22})$. We tested sinusoidal perturbations with different frequencies as well as white noise perturbations and they all converge to the same steady states. We do not expect our solutions depend on initial conditions.

We used a grid spacing of $h = 0.05$ and tests with finer space grid ($h = 0.02$) presented analogous results, suggesting the numerical calculations are accurate. Although in most cases convergence to the steady state is relatively fast and it can be visually inferred, there are cases, specially near the threshold to prey extinction (vertical line of Fig.\,\ref{bifur1}(b)), where convergence can be very slow. We consider a steady state has been reached when the variation of the total population of species 2 at a given time when compared to $100$ units of time later (recall that time has been rescaled by species 1 growth rate $b_1$) is lower than $0.01\%.$  

\end{document}